\documentstyle[12pt,psfig]{article}

\begin{document}
\title{ Is X(3872) a possible candidate of hybrid meson }
\author{Bing An Li\\
Department of Physics and Astronomy, University of Kentucky\\
Lexington, KY 40506, USA}

\maketitle
\begin{abstract}
A study of the possibility of X(3872) as a hybrid state of $c\bar{c}g$ is presented.
The associate productions of X(3872) and $J/\psi$ in hadron collisions and 
flavor Independence of hadrons in $ee^+\rightarrow X(3872)+hadrons$ and
the decays $X(3872)\rightarrow\gamma+hadrons$ are discussed.
It is pointed out that $X(3872)\rightarrow J/\psi+\sigma$ could be a significant 
decay channel.
\end{abstract}
\newpage 
X(3872) has been discovered by the Belle Collaboration[1] in the channel
\[B^{\pm}\rightarrow X(3872)K^{\pm},\;\;X(3872)\rightarrow J/\psi\pi^+\pi^-.\]
The mass of X has been determined to be 
\[m_X=3872.0\pm0.6\pm0.5 MeV.\]
The decay width of X(3872) is less than 2.3 MeV at $90\%$ C.L.[1]. This 
discovery has been 
confirmed by CDF[2], D0[3], and BABAR[4] Collaborations. Different theoretical
conjectures have been proposed for X(3872): a charmonium state[5], a 
molecular state[6],
and a hybrid charmonium state[7].
However, there are challenges from experimental measurements. Charmonium state should
have a large radiative transition rate, $X(3872)\rightarrow \chi_{c1}\gamma$, 
which has not been observed by Belle[1]. The molecular state, ${D}\bar{D}^{*}$,
besides neutral state there should be charged molecular states, $X^{\pm}$,
which decay to $J/\psi\pi^{\pm}\pi^0$. Recently, the BABAR Collaboration[8] has 
reported the measurements
\[B(\bar{B}^0/B^0\rightarrow X^{\pm}K^{\mp}, X^{\pm}\rightarrow J/\psi \pi^{\pm}\pi^0)
<5.8\times10^{-6}\]
\[B(B^{\pm}\rightarrow X^{\pm}K^0_S, X^{\pm}\rightarrow J/\psi \pi^{\pm}\pi^0)
<11\times10^{-6}\]
at $90\%$ level. Charged $X^{\pm}$ have not been found. 

On the other hand, 
the BABAR results
implicate if $\pi^+\pi^-$ in the decay $X(3872)\rightarrow J/\psi \pi^+\pi^-$ comes
from a isovector, $\rho^0$, then 
\[B(B\rightarrow X^\pm K, X^\pm\rightarrow J/\psi \rho^\pm)=2 
B(B\rightarrow X K, X\rightarrow J/\psi \rho^0)=2(1.28\pm0.41)\times10^{-5}\]
which is much greater than BABAR's experimental results[8]. 
Therefore, X(3872) is not a isovector. 
To confirm
the isoscalar state of $\pi^+\pi^-$ it is necessary to search for $X(3872)\rightarrow
J/\psi\pi^0\pi^0$ and to see whether
\[B(B\rightarrow X(3872) K, X(3872)\rightarrow J/\psi\pi^0\pi^0)=0.5\times
B(B\rightarrow X(3872) K, X(3872)\rightarrow J/\psi\pi^+\pi^-)\]
\[=0.5\times
(1.28\pm0.41)\times10^{-5}.\]

In this letter possible tests of X(3872) as a $c\bar{c}g$ hybrid state are
studied. According to Ref.[7], if X(3872) is a $c\bar{c}g$ hybrid state the
dominant strong decay mode is $X(3872)\rightarrow J/\psi+gg, gg\rightarrow
\pi^+\pi^-$. The G-parity of X(3872) is -1. Obviously, in the chiral limit $\rho$ meson
cannot be produced by two gluons. Therefore, BABAR's measurements[8] are supported
by the mechanism of $X(3872)\rightarrow J/\psi+gg$. The decay mode $X(3872)\rightarrow
J/\psi+\sigma$ is allowed by this mechanism. The measurement of $X(3872)\rightarrow
\sigma$ is significant.
Recently, BELLE[9] has reported a new measurement
\[B\rightarrow X(3872)K, X\rightarrow J/\psi\pi^+\pi^-\pi^0.\]
"$X\rightarrow J/\psi\omega$ occur via virtual $\omega$'s and the 3$\pi$ masses 
cluster at the kinematic limit"[9] and it has been determined
\[\Gamma(X\rightarrow J/\psi\omega)/\Gamma(X\rightarrow J/\psi\pi^+\pi^-)
=0.8\pm0.3\pm0.1.\]
Within the experimental errors the rates of the two decay modes are at the same 
order of magnitude. If $X\rightarrow J/\psi3\pi$ is a strong decay mode, the
G-parity of X(3872) is positive, which is contrary to \(G=-1\) determined by
$X\rightarrow J/\psi+2\pi$. It is possible that 
one of these two decay modes violates the conservation
of G-parity.  
It is well known that G-parity violation is caused by either $m_d-m_u$ or 
electromagnetic interactions. However, the effects of G-parity violation is only 
about
few percent and it is about 2$\sigma$ deviation from the central experimental
value[9]. In order to solve the puzzle of G-parity besides to improve the 
experimental accuracy,
we propose to measure another decay mode $B\rightarrow X(3872)K,
X(3872)\rightarrow J/\psi\pi^0\gamma$. If 
$X(3872)\rightarrow J/\psi\pi^+\pi^-\pi^0$ is indeed the process
$X(3872)\rightarrow J/\psi\omega, \omega\rightarrow\pi^+\pi^-\pi^0$[9], then it
should be expected
\[\Gamma(X\rightarrow J/\psi\omega, \omega\rightarrow\pi^0\gamma)\sim0.1
\Gamma(X\rightarrow J/\psi\omega, \omega\rightarrow 3\pi)=
0.1(0.8\pm0.3\pm0.1)
\Gamma(X\rightarrow J/\psi\pi^+\pi^-).\]
As a matter of fact the strong decays of X(3872) are allowed. Therefore,
the issue of the G-parity mentioned above is not for hybrid state only and
it is a very general problem.

The hybrid charmonium has been studied by various theoretical 
approaches[10,7]. 
The mass of $c\bar{c}g$ is at about 4GeV.
The main decay channel of $c\bar{c}g$ is $J/\psi gg$. Therefore,
the decay width is narrow. As suggested in Refs.[7] $c\bar{c}g$ can be found in 
B decays. One of the signature decay channel of $c\bar{c}g$ is $J/\psi \pi^+\pi^-$[7].
All these three properties of $c\bar{c}g$ agree with X(3872). 
 
In this letter we further explore the possibility that X(3872) is a 
hybrid state of $c\bar{c}g$. We propose to study more processes in which X(3872) is
involved to collect more information.
 
X(3872) has not been found by CLLEO's $\gamma\gamma$ fusion experiment[11],
$\gamma\gamma\rightarrow X\rightarrow J/\psi\pi^+\pi^-$ 
\[(2J+1)\Gamma_{\gamma\gamma}B(X\rightarrow J/\psi\pi^+\pi^-)<16.7 eV.\]
If X(3872) is a hybrid of $c\bar{c}g$ and the dominant decay channel is $J/\psi gg$.
Qualitatively there should be strong suppression on the coupling between $\gamma\gamma$
and $J/\psi gg$. On the other hand, 
for the channel $J/\psi gg\rightarrow J/\psi \pi^+\pi^-$(there is already
suppression from the hadronization of two gluons) $J/\psi$ can couple to a photon and 
because Bose statistics isoscalar $\pi\pi$ cannot be spin-1 and a real photon cannot
couple to the pion pair of the coupling $XJ/\psi\pi^+\pi^-$. 
Of course, $J/\psi$ and charged pion can couple to a photon respectively. Therefore,
$\gamma\gamma\rightarrow X(3872)+\pi^+\pi^-$ is allowed.

X(3872) has been found by CDF[2] and D0[3] in $p\bar{p}$ annihilations
\(p\bar{p}\rightarrow X+...\).
If X(3872) is a $c\bar{c}g$ hybrid state, the dominant coupling is $X(3872)J/\psi gg$.
The associate production of X(3872) and $J/\psi$ should be expected in hadron 
collisions. This associate production  
can be understood by the mechanism of fusion of two gluons: $gg\rightarrow X+J/\psi$.
In the parton model the cross section of the associate production 
of X(3872) and $J/\psi$ is written as
\begin{equation}
\sigma(h_1+h_2\rightarrow X+J/\psi+...)=\int^1_{x_{1min}}\int^1_{x_{2min}}dx_1 dx_2
\{G^{h_1}_{g_1}(x_1)G^{h_2}_{g_2}(x_2)+G^{h_1}_{g_2}(x_1)G^{h_2}_{g_1}(x_2)\}
\sigma(g_1+g_2\rightarrow X+J/\psi),
\end{equation} 
where $G^h_g(x)$ is the gluon distribution function of hadron and
\[x_{1min}=(m_1+m_2)/S,\]
\[x_{2min}=(m_1+m_2)/(x_1 S),\]
$m_1, m_2$ are the masses of the two hadrons, \(S=(p_1+p_2)^2\). 
In this process X(3872) decays to $J/\psi \pi\pi$. Therefore, two $J/\psi$'s are 
produced in hadron collisions.  
D0 has 
measured the distribution of the transverse momentum of X(3872)[3]. Model
is needed to calculate $p_\perp$ distribution.
Using the parton model(1), it is obtained
\begin{eqnarray}
\frac{d^2\sigma}{dp^2_\perp}(h_1+h_2\rightarrow X+J/\psi+...)&=&\frac{1}{16S\pi^2}
\int^1_{x_{1min}}\int^1_{x'_{2min}}dx_1 dX_2
\{G^{h_1}_{g_1}(x_1)G^{h_2}_{g_2}(x_2)+G^{h_1}_{g_2}(x_1)G^{h_2}_{g_1}(x_2)\}
\nonumber \\
&&\frac{1}{x_1 x_2}\frac{1}{p_\parallel E_J+p_\parallel(J)E_X}|T|^2,
\end{eqnarray}
where 
$p_\perp$ is the transverse momentum of X, $p_\parallel$ and $p_\parallel(J)$ are
the components of the momenta of X and $J/\psi$ along the direction of the momentum 
of $h_1$ respectively,
\[x'_{2min}={1\over x_1 S}\{(m^2_X+p^2_\perp)^{{1\over2}}+
(m^2_J+p^2_\perp)^{{1\over2}}\},\] 
\[p_\parallel=(x_1-x_2){\sqrt{S}\over2}-p_\parallel(J),\]
$p_\parallel(J)$ can be determined by energy conservation, T is the matrix element
of $gg\rightarrow X+J/\psi$. Obviously, the distribution of the transverse momentum 
of X(3872) can be used to determine the spin of X(3872). The determination of the spin
of X(3872) is beyond the scope of this paper. 

Now we study the electromagnetic decays and productions of X(3872).
It is known that $J/\psi$ can couple to a photon. A mechanism of the VMD[12] of 
$J/\psi$
is expressed as
\begin{equation}
{\cal L}_{J/\psi\gamma}=eg_J\{-{1\over2}F^{\mu\nu}(\partial_\mu J_\nu-\partial_\nu 
J_\mu)+A^\mu j_\mu\}
,
\end{equation}
where $j_\mu$ is a hadronic vector current.  
According to Ref.[12], the VDM(3) is equivalent to the substitution
\begin{equation}
(J/\psi)_\mu\rightarrow eg_J A_\mu.
\end{equation}
Using this substitution(4) in the Lagrangian in which there is the field of $J/\psi$,
the current $j_\mu$ is obtained.  
The constant $g_J$ of Eqs.(3,4) is determined by fitting $J/\psi\rightarrow ee^+$
\[g_J=0.0887.\]
$J/\psi\rightarrow\eta_c+\gamma$ and 
$\eta_c\rightarrow\gamma\gamma$ can be used to test the VMD of $J/\psi$(3,4). 
The amplitudes of  
these two processes are connected by 
the VMD(3,4).
\begin{eqnarray}
\lefteqn{{\cal L}_{J/\psi\rightarrow\eta_c\gamma}=e{A\over m_J}\eta_c
\epsilon_{\mu\nu\alpha\beta}\partial^\mu J^\nu\partial^\alpha A^\beta,}\nonumber \\
&&{\cal L}_{\eta_c\rightarrow\gamma\gamma}=e^2g_J{A\over m_J}
\epsilon_{\mu\nu\alpha\beta}\partial^\mu A^\nu\partial^\alpha A^\beta.
\end{eqnarray}
The parameter A is determined by $\Gamma(J/\psi\rightarrow\eta_c\gamma)$[13] to be
\[A^2=3.06(1\pm0.34).\]
From Eq.(5) it is obtained
\[\Gamma(\eta_c\rightarrow\gamma\gamma)=11.2(1\pm0.34)keV.\]
The data[13] is \(7.44(1\pm0.5)keV\). Theory agrees with data within the experimental 
errors.
It is not the purpose of this letter to present a complete study of the validity 
of the VMD of $J/\psi$
in charm physics. We just use Eqs.(3,4) to discuss the electromagnetic decays 
and productions
of X(3872).

If X(3872) is a $c\bar{c}g$ state, using Eq.(3,4), the amplitude of 
$X\rightarrow \gamma+gg$ is obtained from
$X\rightarrow J/\psi+gg$. The phase space of $X\rightarrow\gamma+gg$ is much larger
than $X\rightarrow J/\psi+gg$. In the radiative decays of X 
the light hadrons are produced
through the process $gg\rightarrow hadrons$. In the limit $m_q\rightarrow 0$ 
($m_q$ is light current quark mass) flavor independence should be expected in
$X\rightarrow\gamma+gg, gg\rightarrow \sigma, 2\pi, KK, \eta', p\bar{p}...$. 
The main component of $\eta$ is SU(3) octet the rate of $gg\rightarrow\eta$
should be small. As a matter of fact BABAR[14] has reported a search for X(3872) in
$B\rightarrow X(3872)K, X(3872)\rightarrow J/\psi\eta$ and only an upper
limit has been determined
\[B(B\rightarrow X(3872)K\rightarrow J/\psi\eta K)<7.7\times10^{-6}\]
at $90\%$ level.  
From flavor independence of $gg\rightarrow hadrons$ we could expect
\begin{equation}
B(X\rightarrow\gamma\pi\pi)/B(X\rightarrow\gamma KK)\sim 1
\end{equation}
\begin{equation}
B(X\rightarrow\gamma\eta)/B(X\rightarrow\gamma\eta')\sim \frac{sin^2\theta}
{cos^2\theta}=0.13,
\end{equation}
where $\theta$ is the mixing angle of $\eta$ and $\eta'$. Here \(\theta=-20^0\) is taken
, which is obtained from the fit of the decay rates of $\eta\rightarrow\gamma\gamma$ 
and $\eta'
\rightarrow\gamma\gamma$[15].
Flavor independence of the radiative decays of X(3872) is a very important feature of
the hybrid state $c\bar{c}g$.

The flavor independence of the $c\bar{c}g$ state can be tested in the production of 
X(3872) in $ee^+\rightarrow X(3872)+gg, gg\rightarrow \sigma, 2\pi, KK, \eta', 
p\bar{p},...$.
According to the VMD of $J/\psi$, a photon can couple to $Xgg$ directly and to $J/\psi$
first and $J/\psi$ couples to $Xgg$.  
The cross 
section of $ee^+\rightarrow X+gg, gg\rightarrow hadrons$ is written as
\begin{equation}
\sigma(ee^+\rightarrow X+hadrons)=3\pi\alpha (g_J)^2\frac{\sqrt{q^2}}{q^4}
\frac{m^4_J+q^2\Gamma^2_J(q^2)}{(q^2-m^2_J)^2+q^2\Gamma^2_J(q^2)}\Gamma
(J/\psi\rightarrow X+Hadrons),
\end{equation}
where 
\begin{equation}
\Gamma(J/\psi\rightarrow X+Hadrons)=\frac{1}{3(2\pi)^3(2\pi)^{3n}}\int d^3p_X
\Pi_i d^3p_i (2\pi)^4\delta^4(q-p_x-p_f)|<f X|{\cal L}|J(q)>|^2,
\end{equation}
q is the total momentum of $ee^+$, $p_f$ is the total momentum of hadrons.
The matrix element $<f X|{\cal L}|J(q)>$ is related to the process 
$X\rightarrow J/\psi gg, gg\rightarrow hadrons$. Model is needed to calculate it. 
We emphasize to use these processes to see whether there is flavor independence
in $ee^+\rightarrow X+f$ where f could be $\sigma, 2\pi, KK, \eta', p\bar{p}...$
\begin{equation}
\sigma(ee^+\rightarrow X\pi\pi)/\sigma(ee^+\rightarrow X KK)\sim 1
\end{equation}
\begin{equation}
\sigma(ee^+\rightarrow X\eta)/\sigma(ee^+\rightarrow X\eta')\sim \frac{sin^2\theta}
{cos^2\theta}=0.13,
\end{equation}
We can estimate the order of magnitude of the cross section of $ee^+\rightarrow 
X(3872)+hadrons$. At the total energy of BEPC, 4.4GeV, the only decay channel is 
$X(3872)+\pi^+\pi^-(\sigma)$. If taking $\Gamma(J/\psi\rightarrow X(3872)+\pi^+\pi^-)\sim 1 MeV$
\[\sigma(ee^+\rightarrow X(3872)+\pi^+\pi^-)\sim 2.3\times10^{-3}nb.\]
If the energy of $ee^+$ is about 10 GeV, more decay channels are open and the
phase space is much larger. If taking $\Gamma(J/\psi\rightarrow X(3872)+hadrons)\sim
1GeV$
\[\sigma(ee^+\rightarrow X(3872)+hadrons)\sim 6\times 10^{-2}nb.\] 

If the energy of photon is above 4 GeV(Jlab for example) then the photoproduction
of X(3872) is another interesting process
\[\gamma+p\rightarrow X(3872)+p,\]
where the VMD of $J/\psi$(3,4) can be used to determine the coupling between photon and 
X+hadrons. In this process the interactions between X(3872) 
and proton
are through $\sigma$, $\pi\pi$, KK, $\eta'$ .... There is also 
\[\gamma+p\rightarrow X(3872)+\pi^+ +n,\;\;\gamma+p\rightarrow X(3872)+K^+ +\Lambda(
\Sigma^0),\;\;\gamma+p\rightarrow X(3872)+K^0+\Sigma^+....\]
There are charged pion and kaon exchange respectively. 

In summary, the associate productions of X(3872) 
and $J/\psi$ in hadron collisions, $X(3872)\rightarrow\gamma+light hadrons$, 
and $ee^+\rightarrow X(3872)+light hadrons$ are proposed to study the nature 
of X(3872). 
If X(3872) is a hybrid
state $c\bar{c}g$ the couplings
$X(3872)\rightarrow J/\psi+gg$ and 
$X(3872)\rightarrow\gamma+gg$ are expected. 
There is flavor independence for the  hadrons produced by two 
gluons, which can be tested. The photoproductions of X(3872) are worth an attention.
$X(3872)\rightarrow J/\psi+\sigma$ is a significant decay channel.

This work is supported by a DOE grant.


\begin{thebibliography}{100}
\bibitem{} S.-K.Choi et al., Belle Collaboration, Phys.Rev.Lett.{\bf 91},262001-1
(2003).
\bibitem{} D.Acosta et al., CDF Collaboration, Phys.Rev.Lett.{\bf 93}, 072001(2004).
\bibitem{} V.M.Abazov et al., D0 Collaboration, hep-ex/0405004.
\bibitem{} B.Aubert et al., BaBar Collaboration, hep-ex/0406022.
\bibitem{} E.Eichten, K.Lane, and C.Quigg, Phys.Rev.Lett.,{\bf 89},162002(2002)
and Phys.Rev.{\bf D69},094019(2004);
T.Barnes and S.Godfrey, Phys.Rev {\bf D69},054008(2004).
\bibitem{} N.Torqvist, Phys.Lett.{\bf B590},209(2004);M.B.Voloshin, Phys.Lett.
{\bf B579},316(2004); F.Close and P.Page, Phys.Lett.{\bf B578},119(2004); C.Y.Wong, 
Phys.Rev.{\bf C69},055202(2004);E.Braaten and M.Kusunoki, Phy.Rev.{\bf D69},
074004(2004);E.Swanson, Phys.Lett.{\bf B588},189(2004).
\bibitem{} F.E.Close et al., Phys.Rev.{\bf D57},5653(1998); G.Chiladze, A,Falk, 
and A.Petrov, Phys.Rev.{\bf D58},034013(1998); F.Close and S.Godfrey, 
Phy.Lett.{\bf B574},210(2003).
\bibitem{} B.Aubert et al., BABAR Collaboration, hep-ex/0408083.
\bibitem{} F.Fang, Belle Collaboration, talk presented at $32^{nd}$ International 
Conference on High Energy Physics, 8/16-8/22, 2004, Beijing, China.
\bibitem{} R.Giles and S.Tye, Phys.Rev.{\bf D16},1079(1977); S.Peratonis and 
C. Michael, Nucl. Phys.{\bf B347},854(1990); P.Lacock et al.,
Phys.Lett.{\bf B401},308(1997); C.Bernard et al., Phys.Rev.{\bf D56},7039(1997);
K.Juge, J.Kuti, and C.morningstar, hep-lat/9709131; hep-lat/9709132; hep-lat/9711451;
N.Isgur and J.Paton, Phys.Rev.{\bf D31},2910(1985);J.Merlin and J.Paton, ibid. 
{\bf D35},1668(1987); T.Barnes ,F.Close, and E.Swanson, ibid.,{\bf D52},5242(1995);
P.Hasenfratz et al., Phys.Lett.{\bf B95},299(1980); S.Ono, Z.Phys. 
{\bf C26},307(1984);J.Latorre et al., Phys.Lett.{B147},169(1984).
\bibitem{} Z.Metreveli et al., CLEO Collaboration, hep-ex/0408057.
\bibitem{} J.Sakurai, Currents and Mesons, Univ. of Chicago Press, Chicago, 1969.
\bibitem{} Particle Data Group, Phys.Lett.{\bf B592}, 1(2004).
\bibitem{} B.Aubert et al., BaBar Collaboration, hep-ex/0402025.
\bibitem{} B.A.Li, Phys.Rev.{\bf D52},5184(1995).
\end{thebibliography}
\end{document}